\documentclass[letter]{aa}
\usepackage{natbib}
\usepackage{graphicx}
\usepackage{txfonts}
\def\arcsec{\hbox{$^{\prime\prime}$}}

\def\degree{^\circ}
\begin{document}
   \title{Spectroscopic evidence for helicity in explosive events}
   \author{W.~Curdt\inst{1}, H.~Tian\inst{2}}
   \institute{Max-Planck-Institut f\"ur Sonnensystemforschung (MPS),
   Max-Planck-Str.2, 37191 Katlenburg-Lindau, Germany\\
   \email{curdt@mps.mpg.de}
\and High Altitude Observatory, National Center for Atmospheric Research, Boulder, CO, USA.
             }
   \date{Received May 17, 2011; accepted June 27, 2011}

  \abstract
  {
  }
   {We report spectroscopic observations in support of a novel view of transition region
   explosive events,
   observations that lend empirical evidence that at least in some cases explosive events
   may be nothing else than spinning narrow spicule-like structures.}
   {
   Our spectra of textbook explosive events with simultaneous
   Doppler flow of a red and of a blue component are extreme cases of
   high spectroscopic velocities that lack apparent motion, to be expected if interpreted as
   a pair of collimated, linearly moving jets. The awareness of
   this conflict led us to the alternate interpretation of redshift and blueshift
   as spinning motion of a small plasma volume. In contrast to the bidirectional jet
   scenario, a small volume of spinning plasma would be fully compatible
   with the observation of flows without detectable apparent motion. We
   suspect that these small volumes could be spicule-like structures and try to
   find evidence. We show observations of helical motion in macrospicules and
   argue that these features -- if scaled down to a radius comparable to the
   slit size of a spectrometer --  should have a spectroscopic signature similar
   to that observed in explosive events, while not easily detectable by imagers.
   Despite of this difficulty, evidence of helicity in spicules has been reported in the
   literature. This inspired us to the new insight that the same narrow spinning
   structures may be the drivers in both cases, structures that imagers observe as
   spicules and that in spectrometers cross the slit and are seen as explosive events.
    }
   {
   We arrive at a concept that supports the idea that explosive events and
   spicules are different manifestations of the same helicity driven scenario.
   In contrast to the conventional view of explosive events as linear bidirectional
   jets, that are triggerred by a reconnection event in the transition region,
   this new interpretation is compatible with the observational results. Consequently,
   in such a case, a photospheric or subphotosperic trigger has to be assumed.
   }
   {We suggest that explosive events / spicules are to be compared to the unwinding
   of a loaded torsional spring.}

   \maketitle
%
%________________________________________________________________

\section{Introduction}

Explosive events (EEs) are characterized as short-lived, small-scaled incidents
of rapid plasma acceleration to typically 50~km/s\, to \,150~km/s and
sometimes even higher velocities in both directions. In spectroscopic data
EEs are easily detected by the redshift and blueshift of the observed transition region (TR) line.
The terminology 'explosive event' has first been introduced by \citet{Dere84}
based on the analysis of high-resolution spectra of TR emission lines
obtained by the HRTS instrument on {\it Spacelab},
but it turned out that this term is quite debatable \citep[e.g.,][]{Dere89}
and may be misleading.% if compared to a typical explosion.
With the advent of {\it SoHO}-SUMER \citep{Wilhelm95} a revival of this field of research
started. Typical EEs were found to be short-lived (60~s to 200~s), small-scale (1500~km to 2500~km),
high-velocity ($\pm$50~km/s to $\pm$150~km/s) flows that occur very frequently, sometimes in bursts.
\citet{Teriaca04} estimated an average size of 1800~km, a birth rate of
2500~s$^{-1}$, and 30,000 events at any one time on the entire Sun. In the classical
view EEs are seen as bi-directional jets that are generated by a
Petschek-type reconnection event \citep{Innes97} high up in the TR with a
collimated upflowing component -- blueshifted in TR emission -- and with a
downflowing, redshifted component at some angle to the line-of-sight (LOS). Statistically,
the blueshift dominates the redshift in magnitude \citep[e.g.,][]{Innes97,Madjarska02,Zhang10},
a fact that is explained
by the deceleration of the downflowing material hitting denser atmospheric layers.
%A wealth of literature has been published since then.
\citet{Ning04} found that EEs tend to cluster near regions of evolving network fields and speculated that
the periodicity of 3\,-\,5 minutes found in EE bursts may be related to subsurface phenomena.
An overview of the immense amount of work on small-scale dynamical events and related cross-references
are found in the review of \citet{Innes04}.
% Recently, \citet{Zhang10} investigated the relationship of EEs and the self-reversal of the Lyman-$\beta$ line.

From the very beginning, however, there has been the problem that apparent motion
as the result of such high-velocity events has not been observed by
imaging instruments. \citet{Dere89} already noted {\it '... the lack of detectable apparent motions
of such high-velocity events'}. \citet{Innes04} again mention this conflict that is still unsolved.
It is the central point of our work to spur a discussion on this discrepancy.

It is intriguing to see that other solar phenomena exist with very similar characteristics
in terms of size, duration, temperature, occurrence rate, light curves, or repeatability.
\citet{Madjarska03} tried to establish a relationship with limb phenomena,
that are observed with a different geometry, and suggested that blinkers
(as observed by CDS) are the on-disk signature of spicules.
While in this article they still assume that blinkers and EEs are
two separate phenomena not directly related or triggering each other, they
later \citep{Madjarska09} state that {\it 'the division of small-scale
transient events into a number of different subgroups, for instance
EEs, blinkers, spicules, surges or just brightenings, is ambiguous'}. Also
\citet{Wilhelm00} argues that there seems to be no obvious distinction between macrospicules
and other spicules, apart from the fact that macrospicules are restricted to coronal hole locations.
With this proposition, \citet{Madjarska06} conclude that blinkers,
EEs and macrospicules are indeed identical phenomena that are
observed with different instruments and with a different geometry.
They, however, still assume flows of rising and falling plasma.

%\citet{Innes98} present a coronal observation that they explain with a bi-directional jet
%configuration. However, assuming a spinning plasmoid could also explain this
%observation without any problem.
\citet{Innes99} present a 2D-reconnection model for EEs, yet they already
mention the possibility of an alternate interpretation of spinning plasma.
This ambiguity of a configuration as bi-directional jet or as a spinning volume
of gas is explicitely mentioned by \citet{Innes01}, but still unsolved.
The dynamical events presented in the following sections clearly favour the
latter explanation.

Helicity is often observed in large-scale events like coronal mass ejections, and
is also found in small-scale phenomena like coronal bright points and X-ray jets \citep[e.g.,][]{Shen11,Liu11}.
\citet{Tian08} found that the Doppler shift pattern of a coronal bright point (BP)
gradually varies with height, suggesting that the magnetic loops associated
with the BP are twisted or in helical form.
Recently \citet{Kamio10} communicated the observation of an X-ray jet with helical motion
at TR temperatures observed by both spectrometers {\it Hinode}-EIS and {\it SoHO}-SUMER.
\citet{Nistico09} report in their survey of {\it STEREO}-EUVI jets that 31 out of 79
events exhibit helical motion, and further mention the possibility that
the rest were very narrow so that possibly the twist could not be resolved. This
notion has recently been supported by \citet{Sterling10}, who suggest that macroscopic
coronal jets can be scaled down to spicule-size features.

We present a case study of two EEs to demonstrate the conflict
with the standard bi-directional jet model and to stress
the enigmatic discrepancy of lacking apparent motion. As a solution of this
conflict we suggest -- backed up by observational evidence for helicity in
spicules -- that a spinning motion may be the source of EEs.
Combining the hypothetical concept of spinning spicules with the observation of quasi-stationary
Doppler-flow in EEs could be the solution of the conflict.

%\section{Observations}

\section{Observation of explosive events}

In the period from Nov 12\,-\,19, 2010, SUMER ran a campaign to observe sunspots
in TR emission lines. We report two cases of EEs found outside, but in the neighbourhood of a sunspot on Nov 16
and on Nov 19, referred to as EE1 and EE2.

\begin{figure}[thb]
\centering
\includegraphics[width=8.8cm]{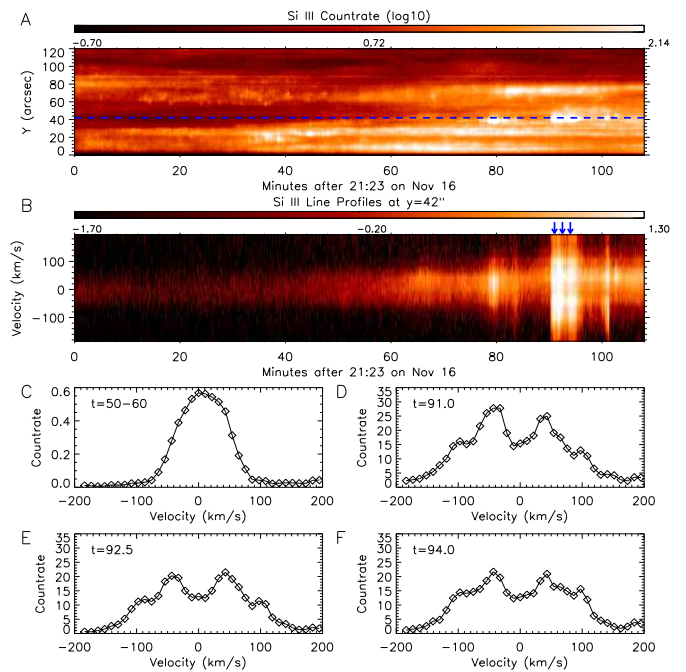}
\caption{
Evolution of profiles of EE1 observed on Nov 16 around 22:55.
The profile of the Si\,{\sc iii} line splits into two components; over several minutes
upflow and downflow stay unchanged within the 0.3\arcsec wide slit.
}
\label{profiles}
\end{figure}

On Nov 16, 2010 SUMER observed the leading sunspot of active region 11124.
The slit of size 0.3\arcsec $\times$ 120\arcsec was placed in such a way that during one hour the
drift by the solar rotation would allow to image the entire spot.
A spectral window around the optically thin emission line of Si\,{\sc iii} at 12.06~nm
%and H\,{\sc i} at 12.16~nm
was read out at a cadence of 10~s. Standard data reduction procedures were applied
to process the data set. In Fig.1 we show the drift scan as $y-t$ plot (top),
the $\lambda-t$ plot (below top) and line profiles in pixel 42 as indicated by the dashed line.
At this location in the plage area very close to the sunspot, a rapid brightness increase
by a factor of $\ge 20$ is observed at time step 91. The pre-event profiles have been
averaged and three more profiles are shown through the event. The timing is indicated by blue arrows.
Interestingly, the spectral line seems to split into two main components that are
symmetrically shifted by 40~km/s towards the red and towards the blue with
additional components at $\pm$100~km/s that are less strong.
The lightcurve is not flat, it has two maxima that are separated by
$\approx$120~s, but the overall shape with four peaks does not change over more than
three minutes.
%The event is also seen in the Lyman-$\alpha$ line,
%although the impact on the line profile is complex and difficult to interpret.

A similar observation with EE2 was completed on Nov 19, 2010, when the instrument
was pointed to the leading sunspot in AR 11126. Again, a stationary double-component
EE with velocities of $\pm$35~km/s is observed from 21:53:14 to 21:57:14 in a plage
location. Similar to the case of EE1, the brightness jumps by a factor of 10 and
the lightcurve is double-peaked (cf., Fig.2). During both events the Sun has rotated by
$\approx$0.4" which is significantly below the spatial resolution of SUMER of 1.5".

\begin{figure}[thb]
\centering
\includegraphics[width=8.8cm]{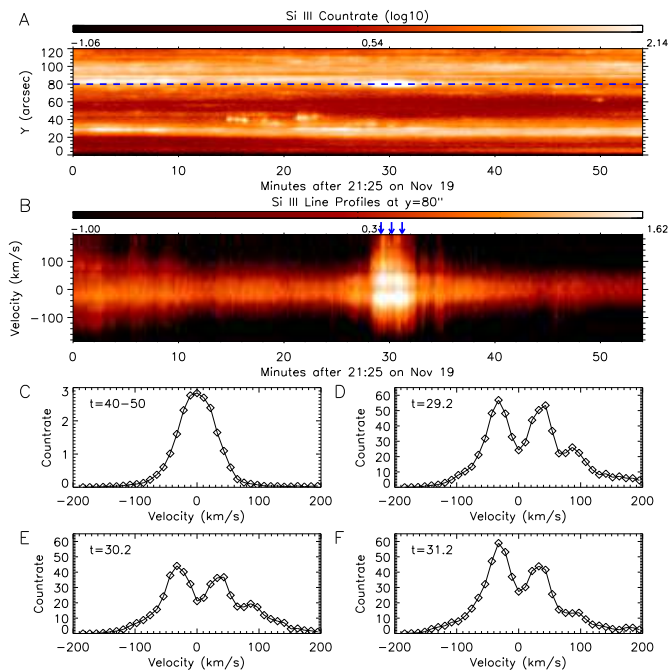}
\caption{
Evolution of profiles of EE2 observed on Nov 19 around 21:55.
Again, the line profile line splits quasi-stationary into two components.
}
\label{jet}
\end{figure}

As a by-product of this study we identified two emission lines that were
also recorded in the Si\,{\sc iii} window and not included in the SUMER spectral
atlas \citep{Curdt01} as S\,{\sc i} lines. In the atlas
%Lyman-$\alpha$ was placed on the attenuator, and consequently
the Si\,{\sc iii} window was recorded on the bare photocathode,
while in this data set we could place it onto the KBr coated photocathode. All S\,{\sc i} lines
at 120.204~nm, 120.261~nm, 120.344~nm, 120.435~nm,120.559~nm, 120.613~nm, 120.704~nm,
120.776~nm, 120.886~nm, 121.122~nm, and 121.018~nm are present in SUMER spectra.

\section{Discussion}

In both events discussed here the Doppler-flow indicates symmetrical flows of
$\approx$40~km/s. It is unclear whether
the 100~km/s components of EE1 are Doppler flows, since they can also be
interpreted as blends by the S\,{\sc i} lines at 120.613~nm and 120.704~nm
(as discussed above). In case of an interpretation as Doppler flow, this would
imply a multi-component event with two sources in the slit area. Because of the
ambiguity however, we do not discuss this issue any further.

The Doppler-flow pattern seen in the line profiles is almost unchanged in
magnitude of the line shift and in the location along the slit.
This excludes any lateral movement exceeding $\approx$500~km along or across the slit
within the 3-minute duration of the event. Within 3~minutes however,
a bi-directional jet moving at 40~km/s should have reached 7,200~km in each
component. This requires that the direction of the jet deviates less than $1\degree$
from the LOS. Such a scenario is very unlikely. It is simply not possible that
in such an event
upflow and downflow stay over minutes stationary within the 0.3\arcsec wide slit.
Also the fact, that no increase in size along the slit is observed is a strong
argument against a linear moving, collimated jet. A similar argument holds for EE2,
that lasts even longer.

The conflict of lacking apparent motion is so evident in the
examples shown here, that we now adopt the alternative flow configuration. If
we assume a spicule-like feature that is as narrow as as the spectrometer
slit and crosses the slit at some angle below $90\degree$, then the redshifted portion
and the blueshifted portion will appear simultaneously in spectroscopic data and
can stay without apparent motion for an extended period of time, exactly as observed in EE1 and in EE2.

The double-peak in the emission of EE1 and EE2 may also be
an effect of the spinning motion, if we assume the repetition of a brightness
maximum after completing a full revolution after $\approx$200~s. Alternatively, the double-peaked
lightcurve may have something to do with the occurrence of double-threaded
jets that have been reported from XRT observations \citep{Kamio10}.
%In all likelyhood the spinning plasma volume also has an upflow component,
%resulting into a spiraling motion.

Motivated by the plausible solution of the old discrepancy we looked for suitable
candidates of solar phenomena as conceivable counterparts for our double-component EEs.
Such candidates could be type II spicules or Rapid Blueshifted
Events (RBEs) \citep{DePontieu09,DePontieu11,Scott09,vdVoort09} since they have very similar
characteristics in terms of velocity, lifetime, size, and repeatability. A
direct proof of helicity in RBEs by imaging instruments has to our knowledge not been
reported yet and may be difficult to achieve. There is, however, indirect
evidence, since rotation in macrospicules -- believed to be bundles of
substructures -- was often observed. On the small side, evidence for helicity in regular spicules
has been reported in literature as already mentioned.
Fig.3 shows several spicules and a macrospicule in a SUMER raster in O\,{\sc v}
obtained on August 18, 1996 in a coronal hole location.
%This raster is also shown in the SUMER atlas (Feldman et al. 2003, page A.162).
The panels show a brightness raster (left) and a dopplergram (right). It is obvious that the macrospicule is
spinning like a bended cylinder, but there is no signature of this motion in
the spectroheliogram. This demonstrates that even in such large structures imagers are principally unable
to observe the rotation, unless finestructures can be resolved.
Similar observations of 'tornados' have been reported by \citet{Pike98} from
{\it SoHO}-CDS data.
We use the observed helicity in a macrospicule -- a much larger feature
than the EEs discussed here -- together with the published
premise of no obvious distinction between macrospicules and other spicules as
support for our argument.

\begin{figure}[thb]
\centering
\includegraphics[width=8.8cm]{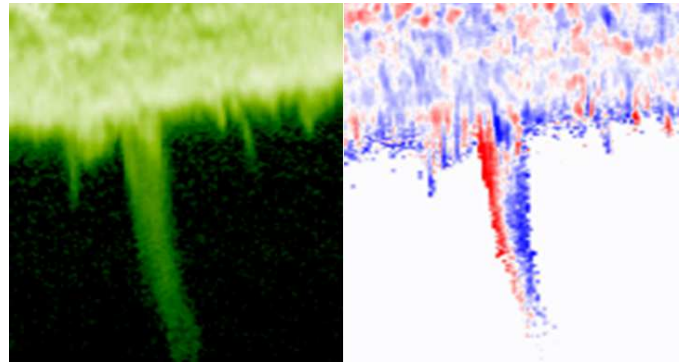}
\caption{
A macrospicule at the solar limb near the South pole is observed in the light
of the O\,{\sc v} line at 62.9 nm, corresponding to 230,000 K; pseudo colours represent radiance (left)
and Doppler motion (right). The Doppler flows are scaled from +30~km/s (red) to
-30~km/s (blue). One half of the plasma ejection moves towards us, the other half away from us;
the spicule swirls like a tornado along a magnetic field line with an Earth-sized diameter rotating at $\pm$30~km/s.
}
\label{spicule}
\end{figure}

The cartoon in Fig.~4 shows typical SUMER line profiles calculated for a spiraling spicule at various
aspect angles. We assume three components of the total radiance, two from the spinning motion with
a tangential velocity of $\pm$60~km/s that contribute with 47.5\% each.
As the third component we assume a faint flow of 100~km/s along the spicule -- typical
for type-II spicules -- that contributes with 5\%. The angle between spicule (or upflow)
and LOS, $\theta$, is set as 0$\degree$, 30$\degree$, 45$\degree$, 90$\degree$, 135$\degree$, 150$\degree$, and 180$\degree$
for the seven cases (the cases with $\theta > 90\degree$ are mirror symmetric to cases (A) to (C) and
not shown in Fig.~4). The spectrum in case (C) is very similar to those
presented here and can quantitatively reproduce our observations.

\begin{figure}[t]
%\centering
\includegraphics[width=8.8cm]{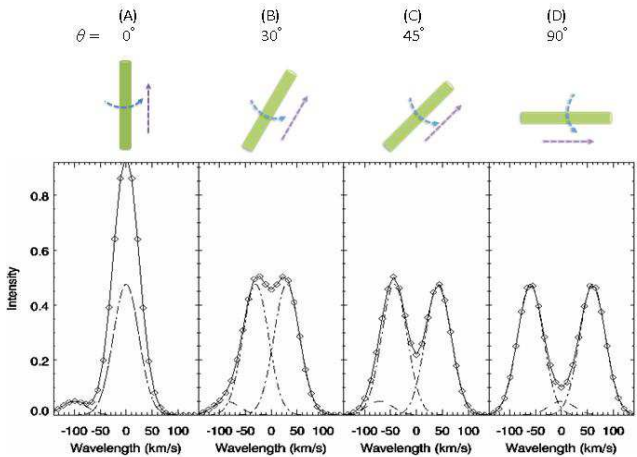}
\caption{Cartoon showing the suggested configuration, blue arrows indicate the spin,
violet arrows the flow direction. We assume three-component emission, red and blue components from the spinning
spicules (dot-dashed, each accounting for 47.5\% of the total
emission), the third component is the faint upflow along spicules (dashed lines, 5\% of the total emission).
The Gaussian width of each component is set to a typical value of 35 km/s.
The spectral resolution is 11~km/s, comparable to a SUMER pixel. The composed emission is
shown as the diamonds connected by solid lines.
}
\label{cartoon}
\end{figure}

In many observations the red and the blue component of the EE are observed together with
a component at rest. We attribute this zero-velocity component to
the background emission of the solar disk that is also visible in optical
thin emission. In our case, however, the foreground
emission of the EEs is so much stronger that it outshines the background.

The fact that in SUMER spectra no EEs are observed in coronal lines is -- besides
the fact that no really good coronal lines for disk observations exist in the
SUMER wavelength range -- not in contradiction to the scenario suggested by
\citet{DePontieu11}, who found that the heated volume is outside the leading edge of the jet.
If heating takes place while the jet propagates and expands, then spectrometers,
in particular slow spectrometers, will have difficulties to observe such a
heating process.

Although none of the AIA channels covers the temperature regime around
46,000 K, the formation temperature of Si\,{\sc iii}, we tried to find whether signatures
of the SUMER EEs are found in {\it SDO}-AIA images, but we could not find any.
This is not too surprising for a feature that
is in contrast to the macrospicule shown in Fig.~3 as narrow as the size
of the SUMER slit, i.e. below the AIA spatial resolution of 1.2\arcsec.

The radius of the macrospicule in Fig.3, $r$, is measured as 5500~km. At the
periphery, the Doppler flow is \mbox{$v=\pm$~30~km/s.} This allows to determine the centripetal
acceleration $a_c=v^2/r$. The value $a_c=0.18$~km/s$^2$ is comparable to the
gravitational acceleration. \citet{Pasachoff68} did a similar exercise for spinning
spicules and arrived at a value of $a_c=1.8$~km/s$^2$. From the examples shown as EE1
and EE2, which are much smaller than the macrospicule we estimate a value of
$a_c=7$~km/s$^2$ assuming $v=\pm$~40 km/s and $r$ as about the slit size of 0.3\arcsec.
Similar values can be expected, if we adopt typical parameters mentioned by
\citet{Teriaca04}, namely a diameter of 2\arcsec and a velocity of 150~km/s. Such violent motions
could contribute to the solar wind acceleration \citep{Pasachoff68}.

\section{Conclusion}

We propose a hypothesis suggesting that disk EEs and limb spicules are the same
phenomenon. However, we assume an alternateive configuration to explain the flows.
It is clear that helicity is behind both manifestations
and should be included to understand the physical nature of EEs.
The assumption of a swirling narrow cylindrical body, rotating while
upflowing, can reproduce observations by both imagers and spectroscopes
which can explain the discrepancy between spectroscopic motion and apparent motion.
Also, the statistical blueshift dominance of EEs would be an obvious consequence
in such a scenario  (see Fig.4, case c). Although
our observations do not strictly exclude the possibility of bi-directional
jets, there are good reasons to assume that EEs are indeed the spectroscopic
signature of spinning type II spicules crossing the spectrometer slit in many cases.
Even more, there seems to be neither an obvious distinction between macrospicules and
microspicules nor between blinkers and EEs.

The swirling component is normally not detectable by filtergraph instruments, but adds considerably to
the energy released by the apparent motion that is detected by such instruments.
We note that the scenario of spinning spicule-like features that are rooted in the photosphere
requires photospheric or even subsurface sources which is not compatible with the
model of reconnection events in the TR. This aspect may require the distinction of
different types of EEs and calls for more systematic work.
We speculate that the helicity may be related to global helicity as generated by the differential
rotation.
%that continually winds up magnetic fieldlines until they are suddenly released by a trigger.
Alternatively, local reconnection due to the subsurface turbulence in twisted flux tubes as
discussed in MHD models could be a possible driving mechanism. In this context, it would be worthwhile
to study the chirality of the events and look for different preferences in both hemispheres.
These hypotheses are, however, not supported by our data and beyond the scope of this communication.
%As a general remark we note that the concept of a spiraling motion inherently includes a periodic
%component, that invokes a dualistic view on waves and oscillations and on jets and outflows.

The IRIS mission -- providing fast spectroscopic capabilities complemented by
a chromospheric imager -- will be an ideal platform for systematic statistical analyses
of geometrical effects and their imprints on the center-to-limb variation
of red-blue tilt, red-blue asymmetry, birth rate, and mean velocity of EEs and
also assessing the dominance of helicity in EEs in a quantitative manner.

\begin{acknowledgements}
The SUMER project is financially supported by DLR, CNES, NASA, and the ESA PRODEX
Programme. SUMER is part of {\it SoHO} of ESA and NASA.
H.T. is supported by the ASP Postdoctoral Fellowship Program of NCAR.
The National Center for Atmospheric Research is sponsored by the National Science Foundation.
We thank I.E. Dammasch, who spotted the beautiful macrospicule for the SUMER image galleries.
We also thank B.N. Dvivedi for critical comments that helped to improve the
clarity of this communication.

\end{acknowledgements}
%\newpage

\end{document}